\documentclass[%
 reprint,
 amsmath,amssymb,
 aps,
]{revtex4-2}

\usepackage{graphicx}% Include figure files
\usepackage{dcolumn}% Align table columns on decimal point
\usepackage{bm}% bold math
\begin{document}

\preprint{APS/123-QED}

\title{Flow Of Information In a Mechanically Quenched Confined Flock}% Force line breaks with \\
\thanks{Correspondence to}%
%~~~~~~~~~~~~~~~~~~~~~~~~~~~~~~~~~~~~~~~~~~~~
\author{Md. Samsuzzaman$^1$, Mohammad Hasanuzzaman$^{2,4}$, Ahmed Sayeed$^1$, and Arnab Saha$^3$}
\affiliation{
\parbox{\linewidth}{$^1$Department of Physics, Savitribai Phule Pune University, Pune 411007, India \\$^2$ICTP – International Centre for Theoretical Physics, Strada Costiera, 11, 34151, Trieste, Italy\\$^3$Department of Physics, University Of Calcutta, 92 Acharya Prafulla Chandra Road, Kolkata-700009, India\\$^4$CINECA High-Performance Computing Department, Casalecchio di Reno, 40033 Bologna, Italy\\
}
}
 \email{sahaarn@gmail.com}
 \email{samsuzz@gmail.com}

%~~~~~~~~~~~~~~~~~~~~~~~~~~~~~~~~~~~~~~~~~~~~~

%\author{Ahmed Sayeed}
% \homepage{}
%\affiliation{
%\affiliation{
%\author{Arnab Saha}
%\affiliation{%


\date{\today}% It is always \today, today,

\begin{abstract}
%Active particles communicate and transfer information to one another for a variety of reasons, whether it might be for foraging food, migration, or escaping threats and obstacles. They do so by interacting with each other and therefore it will be an obvious reason to make use of statistical tools such as information theory to analyze the behavior of active particle clusters when they encounter any change in their surroundings. Here, we introduce an external perturbation by quenching the trap boundary to a perfectly hexatic cluster and measure the information flow between them using transfer entropy. We also found out that this information propagation is ballistic.   

 Living entities in a group communicate and transfer information to one another for a variety of reasons. It might be for foraging food, migration, or escaping threats and obstacles, etc. They do so by interacting with each other and also with the environment. The tools from statistical mechanics and information theory can be useful to analyze the flow of information among the living entities modelled as active (i.e. self-propelling) particles. Here we consider the active particles confined in a circular trap. The self-organisation of the particles crucially depends on whether the trap boundary is soft or hard. We quench the trap boundary from soft to hard instantaneously. After the mechanical quench, the particles suddenly find themselves in a hard potential. The self-organised cluster of the active particles, which was stable when the boundary was soft, becomes unstable. The cluster undergoes extreme deformation after the quench to find another stable configuration suitable for the hard potential. Together with the structural relaxation, information regarding the quench also flows throughout the deforming cluster. Here, we quantify the flow of information by computing local transfer entropy. We find that the flow spans the whole cluster, propagating ballistically.  
\end{abstract}

\maketitle{}

%\tableofcontents

\section{\label{sec:level1}Introduction}

Staying in a group offers numerous advantages and disadvantages, both from a social and evolutionary perspective. Advantages include life-security from predatory assaults with collective defense\cite{Meunier2015,siegfried1975}, resource sharing\cite{Pitcher1982}, enhanced learning\cite{sasaki2018}, social support\cite{Hart2021}, increased reproductive success\cite{ward2016sociality}, thermo-regulation\cite{ancel2015}, efficient division of labor\cite{Kolmes1989}, etc. These advantages play a major role in the survival and success of many species. However, while dwelling in groups, animals may suffer in several ways. For example, one may consider the growing size of a group. Typically the size of a group is expected to vary among species based on the local ecological variables that a population or group is subjected to. However, a group may grow so large that the competition for resources within the group becomes fierce and this outweighs the benefits\cite{MacDonald2008}. Another common disadvantage of living in a group is the increasing risk of spreading contagious infection\cite{Charles1992}. However, for many species the advantages of group living outweigh solitary living for various ecological and evolutionary reasons. 

In a group, animals often interact with each other psychologically rather than with physical forces alone\cite{HelbingPRE1995}. Hence to understand the collective as well as individual dynamics of animals in a group, social forces originating from the psychology of the individuals,  should be factored in. They mediate flow of {\it{information}} among the individuals of a group\cite{AliciaPRSB2022}. To develop a detailed understanding of how members in a group interact with each other and often reach a common consensus, we must examine how information flows in and around the group\cite{Couzin2005}. %Even when trying to modify the fundamental model to accommodate field findings, physicists are not particularly interested in animal social interactions; nevertheless, humans or animal crowds behave differently, motivated more by psychology than by physical forces alone\cite{HelbingPRE1995}. 

The flow of information in a flock involves the communication and coordination processes that occur among individuals within the flock \cite{WilliamPNAS2012}. Flocking behavior is commonly observed in various biological systems, such as in birds, fishes, insects etc.\cite{tkavcikprinceton,JamesPNAS2011,CavagnaPNAS2010,TonerPRE1998,VicsekPRL1995,CouzinASB2003}, where individuals can interact with each other to exhibit organized, collective dynamics. Individual member of a flock must continuously gather and utilise data about the group and also from the external environment in order to function as sophisticated, self-organized, dynamical systems. This information can be exchanged through various mechanisms\cite{GeissPRE2022,LevisPRR2020,BalleriniPNAS2008}, including direct physical interactions, visual cues, acoustic signals, or chemical communication \cite{CavagnaPNAS2010,VicsekPRL1995,DanielPNAS2014,AttanasiNature2014}. The flow of information enables individuals to perform various activities in the flock, such as to align their movements with other members, maintain cohesion, and collectively respond to environmental cues or threats\cite{Garud_rao_PNAS}, etc. The complex spatio-temporal pattern formation and the dynamics of information flow in a flock can be intertwined \cite{moussaid2009collective}  and depend on the sensory ecology \cite{williams2023sensory} of the particular species and the environment or both \cite{PeruaniPRL2013,PeruaniEPJ2015}.

Flocking behavior often relies on local interactions\cite{CavagnaPNAS2010} between neighboring individuals, where each individual typically pays attention to a limited number of its closest neighbors and adjusts its behavior based on their movements. Information is exchanged through these local interactions, allowing for alignment and coordination which finally span the whole flock. In some cases, specific individuals within a flock may have leadership roles\cite{WILL2016}, guiding the overall movement and behavior of the group. Information flows from leaders to followers through signals or behavioral changes. The followers respond to the leaders' cues and adjust their actions accordingly. Visual signals\cite{Peruani2020,Peruani2016,Strombom2011}, such as the relative positions and movements of nearby individuals, play an important role in the flow of information. Other sensory modalities, such as sound or chemical signals\cite{Beatriz2022,Nina2016}, may also be important depending on the species. Information flow in a flock may exhibit complex, non-linear dynamics\cite{Mohammad2013,Uzi2017}. Small changes in the behavior or movement of one individual can spread throughout the flock, leading to cascades of coordinated responses that prevent jamming\cite{BenFabry2011}, collective decision-making, and evasive maneuvers in response to threats\cite{siegfried1975}. 

%********************************our work*********

Flocks are considered as active matter in physics where the constituent elements i.e. the individual members consume ambient energy and become active i.e. motile\cite{Ramaswamy_RMP,Bechinger_Lowen_RMP}.  They are inherently non-equilibrium systems. The self-organisation of the {\it{confined}} motile particles crucially depends on the inter-particle interactions as well as on their interaction with the boundary of the confinement.  It has been shown earlier that keeping the inter-particle interactions unaltered, one can simply alter the properties of the confinement to  tune the collective dynamics and thereby the self-organisation of the active particles (e.g. \cite{RanaSamArnabSoftMatter2019}).  Confinement significantly affects their collective dynamics \cite{LeeNJP2013}. For example, confinement can lead to the alignment and ordering of active particles. When confined within narrow channels, the active particles tend to align their motion such that a collective flow emerges\cite{Wioland_2016}. Their interactions and motion can give rise to various cooperative phenomena, such as swarming, swirling, or vortices\cite{LevinePRE2000}. Confinement can also lead to structural phase transformations in active systems\cite{Araujo2023}. It's important to note that the specific effects of the confinement on the self-organisation and collective phenomena of active systems can depend on various factors, including the nature of the constituent active particles, their interactions and the confinement geometry. Experimental and theoretical investigations continue to explore the complex dynamics of confined active systems. Our understanding of these effects is still evolving.  

Here we quantify the flow of information within a simple, flocking model in a circular confinement. The flock is simulated as a collection of self-propelling, soft, circular disc like particles in two dimensions. While moving, the particles try to align with their neighbours. The particle-based model, although simple, can capture important features of real flock \cite{mora2016local,gomez2023fish}. It has been shown earlier\cite{RanaSamArnabSoftMatter2019} that when the boundary of the confinement is soft, the flocking particles form a rotating (about the centre of the circular trap) as well as rolling (about the centre of mass of the particles) cluster with hexagonal order. However, when the boundary is hard the particles form a thin wetting layer along the boundary. Contrary to the case with the soft boundary, there is no positional order in the particles in the layer. Therefore,  with increasing hardness of the boundary, the system of flocking active particles goes through a phase transformation from a hexagonally ordered state to a positionally disordered state and vice versa. 

In this work we perturb the hexagonally ordered cluster of the active particles trapped within the soft confinement by instantaneously quenching the trap boundary to a very hard one. It is expected from \cite{RanaSamArnabSoftMatter2019} that, the mechanical quench will lead to a structural (i.e. the  order-to-disorder) transition \cite{RanaSamArnabSoftMatter2019} within the self-organised cluster of the particles. Similar transition takes place in many biological systems\cite{BuhlScience2006}. However, here we are primarily interested in the flow of information among the particles after the quench takes place. The flow is initiated by the quench and continues to occur among the particles simultaneously with the structural relaxation. In particular we are interested in measuring the time and length scales of the information flow

Measuring information flow is limited by the definition of information itself. It can be defined as two-point correlations among velocity of the particles\cite{cavagna_giardinaPNAS}. However, here we measure the information flow within the cluster using an information-theoretic measure. Typically these measures are model-independent\cite{TimmeNeuro2018,VicenteJCN2011} and do not necessarily require the specific structure of interaction among the particles. However, it requires the phase-space distribution of the particles.

%but rather require some number that denotes the relationship between variables and also the outcome of such analysis is a number that quantifies some relationship inside the data, and not a parameter in a model(such as the strength of self-propulsion). 

The widespread applicability of the information theory is in neuroscience\cite{AlexanderNatureNeuro1999}, where complex interactions among various neural circuits result in the processing and storing of information. Here we will compute  transfer entropy \cite{SchreiberPRL2000} which will be used to quantify the information flow in groups of interactive active particles \cite{CrosatoSI2018,Lord2016,Cliff2017}. It is a non-parametric way of measuring information flow\cite{SchreiberPRL2000} from neighboring agents to a particular target agent at every time step. It takes care of the history of the target as well as its neighbours while computing the information. We shall go into the details about this in later sections of the paper. From the next section onward, we will systematically go through our findings, starting from a description of the model flock and then moving on to a description of the mechanical quench. Subsequently we compute the information flow among the active particles and reveal its fundamental properties such as, typical length and time scales. Finally we conclude by summarising  our findings.

%Finally, for our system(a compact hexatic-ordered cluster), which encounters a physical change, i.e. quenching of the trap, we use Transfer entropy\cite{SchreiberPRL2000}, which is a non-parametric way of measuring time series information flow, to measure the information flow from neighboring agents to a particular target agent at every time step, to detect the perturbation caused and to check if transfer entropy can measure such structural changes. We shall go into more depth about this in a later section of this paper. We will next systematically go over our findings, starting with a description of the model and moving on to a description of the quench event and analysis of the outcomes of the model's simulation.

% It has already been reported that Mutual Information can be used for identifying transitions in complex systems\cite{WicksPRE2007}, (but transfer entropy also considers directionality so we need to mention this).    
  
\subsection{\label{sec:level2}Model}

We consider individual members of a flocking group as particles. To keep computation simple we take N number of such active (i.e. self-propelling) particles in two dimensions (2D). The particles are indexed with $i$ that runs from $1$ to $N$. The position and velocity of $i$-th particle at time $t$ is denoted by ${\bf{x}}_i(t)$ and ${\bf{v}}_i(t)$.  We consider all the particles have same  mass $m$.

To model the dynamics of such a flock, we write the equation of motion (EoM) of an individual ($i$-th) member of the flock considering all the forces acting on the member.  One of the forces will be the {\emph{active or self-propelling force}} generated within the particle itself consuming available resources (e.g. food) from the surrounding. In reality the magnitude of this force $(f_0)$ may have complex dependence on the rate at which the food is assimilated by the individual,  the metabolic rate of the individual etc., which typically vary from individual to individual. Moreover, for a given individual, the force can even become time-dependent. However, for simplicity here we consider the magnitude of the force to be same for all the individuals, and it is independent of time as well. 

The direction of the active force $\hat n_i(t)$ of $i$-th individual  depends on the direction towards which the individual wants to move. However it can differ from the direction along which it actually moves. This can happen due to, for example, the interactions among other group members. The interaction among the group members are often cognitive and highly complex. However here, to maintain the simplicity of our approach we consider Viscek-like aligning interaction \cite{VicsekPRL1995} among the group members. According to the rule of this interaction, at a given time $t$,  the direction of  the active force of $i$-th individual at the next time step $t+\delta t$ (denoted by $\hat n_i(t+\delta t)$) will be the average direction of active forces of the neighbours of  the individual at time $t$.  Therefore, mathematically  one can write the active force at time $t+\delta t$ as,    

\begin{eqnarray}
{\bf F}_i^{\text{avg}}(t+\delta t)=f_0\hat n_i(t+\delta t)=\frac{f_0}{N_i(t)}\sum_{j=1}^{N_i(t)}\hat n_j(t) 
\label{activeforce_1}
\end{eqnarray}
where $N_i(t)$ is the number of neighbours (indexed as $j$) of $i$-th member at time $t$. However, the time evolution of  the active force of an individual should contain a random component. This randomness can occur due to several reasons. For example, while computing the average direction, the individual can make mistakes or due to  environmental disturbances it may not follow the average direction perfectly. To account for the randomness, ${\bf{F}_i^{\text{avg}}}$ should be rotated with a random angle $\theta$ distributed uniformly within $0$ and $2\xi\pi$. Here $\xi$ is a number between $0$ and $1$ that signifies the range of randomness. If $\xi=0$ there is no randomness in calculating the average direction and if $\xi=1$ it is maximally random. Hence with the finite randomness, the active force ${\bf{F}_i^{\text{active}}}(t+\delta t)$  becomes,   

%\begin{eqnarray}
%{\bf F}_i^{\text{active}}(t+\delta t)={\mathcal{R}}(\theta_i)\circ{\bf{F}}_i^{\text{avg}}=\frac{f_0}{N_i(t)}{\mathcal R(\theta_i)}\circ\sum_{j=1}^{N_i(t)}\hat n_j(t) 
%\label{activeforce}
%\end{eqnarray}

\begin{eqnarray}
%\begin{align*}
 {\bf F}_i^{\text{active}}(t+\delta t)&=&{\mathcal{R}}(\theta_i)\circ{\bf{F}}_i^{\text{avg}} \nonumber\\    
 &=&\frac{f_0}{N_i(t)}{\mathcal R(\theta_i)}\circ\sum_{j=1}^{N_i(t)}\hat n_j(t)
%\end{align*}
\label{activeforce_2}
\end{eqnarray}

Here ${\mathcal{R}}(\theta_i)$ is the 2D rotational matrix. In general it is not necessary that at the individual level, the active force will be generated along the same direction where the particle moves with velocity ${\bf v}_i$. However, as we consider a simple set-up here, we assume $\hat n_i=\hat v_i=\frac{{\bf v}_i}{|{\bf v}_i|}$. \\

One may further note that to compute the direction of the active force according to the aforementioned model, an individual member need to rely on its neighbours. How to choose the neighbours to follow is an important issue for an individual and also for the collective dynamics of the group. In a group it is not always true that a member choose only the geometrically closest members as its neighbours to follow. It is often possible for a member to follow the direction(s) of another member(s) far apart. The mechanism of following each other in a group often depends on the complex social dynamics and hierarchy present among the members of the group. However here, we do not consider such complexities. Instead we consider that an individual chooses only the geometrically closest  group members as its neighbours to follow. For this purpose, at a given time point, an individual member of the group considers only those neighbouring member(s) to follow, who are within a circle having radius much smaller than the linear size of the group itself.

Apart from the active force, the members of the group can also attract each other. Such forces often occurs when the members of a group are of the same species\cite{Hogg_1993}. However when they come too close to each other they repel each other due to steric interactions.  Here we consider that the self-propelling particles are spherically symmetric and interact with each other only via the repulsive force which is derived from a continuous potential $V_{wca}$(Weeks-Chandler-Andersen (WCA) interaction \cite{weeks1971role}) given by,
\begin{eqnarray}
V_{\text{wca}}&=&4\epsilon\left(\frac{\sigma^{12}}{r_{ij}^{12}}-\frac{\sigma^{6}}{r_{ij}^6}\right)+\epsilon \phantom{xxx}
{\text{for}}\phantom{x} r_{ij}<2^{1/6}\sigma \nonumber \\
&=& 0 \phantom{xxxxxxxxxxxxxxxx}{\text{elsewhere}}
\label{WCAEquation}
\end{eqnarray} 
Here $\sigma$ is the measure of the individual particle size and it is the smallest length scale involved in the system, $\epsilon$ determines the strength of the repulsive interaction and the distance between two pairs of particles i.e. i-th and j-th particles is $r_{ij}$. Furthermore, it is important to mention that the cutoff within which this repulsion occurs is $r_{int}$, where $r_{int}=2\sigma$ to account for the steric repulsion. We consider the same cut-off to take into account the alignment of active forces among the neighboring particles. Note that here we consider only the two-body interaction. We assume that the three-body or even higher order interactions are less probable. This essentially implies that the group of particles of our concern is in semi-dilute condition. Though WCA interaction is used to simulate passive i.e. non-living microscopic systems \cite{semenov2013electrophoretic}, recently it is being used to model inter-particle interaction potential for many active systems \cite{ni2015tunable}.    

Next, we consider the trap in which the group of the self-propelling particles is confined. There are many examples of confined active systems.  An example can be bacteria or other motile cells in a micro-fluidic channels\cite{zhang2021}. The interaction between the active particle with the confinement can be very complex in general. It can depend on the physical properties of the wall of the confinement (for example whether it is hard or soft\cite{RanaSamArnabSoftMatter2019}). It may depend on whether the wall is hydrophobic or hydrophilic \cite{krasowska2014microorganisms}.  It may also depend on the geometry (e.g., whether it has finite curvature or not \cite{chang2015biofilm}) of the confinement etc. The confining wall can also considerably influence the hydrodynamic interactions (if any) among the active particles. Hence, the self-organisation of confined, wet, active matter, the wall can play a crucial role \cite{thutupalli2018flow}. However, here barring several complexities we consider dry active matter confined in a 2D circular trap introduced by the potential 

\begin{eqnarray}
U({\bf r}_i)=U_0(\tanh(q(r_0-|{\bf r}_i|))+1)
\label{TrapEquation}
\end{eqnarray}    

Here $U_0$ is the depth of the trap. It ensures that all the active particles should always be confined in the trap. The radius of the trap is $r_0$ and $q$ is its hardness parameter which we can vary here. If $q$ is very large, then the trap wall is very steep as well as hard. The particles near such a wall face strong repulsion.  On the other hand, when $q$ is small the trap becomes smooth and soft. The particles close to a soft wall face lesser repulsion than in case of a hard wall. The dimensionless steepness parameter can be defined as 

\begin{eqnarray}
S=q\sigma. 
\end{eqnarray}

Earlier we have shown that by tuning $S$ one can tune the self-organisation of the active particles from a disordered to an ordered state and vice-versa\cite{RanaSamArnabSoftMatter2019}. This non-equilibrium transition will be exploited here to introduce a mechanical quench causing the flow of information within the group of active particles confined within the trap.   
We also note here that the potential has a constant curvature $1/r_0$.  The force ${\bf{F}}_i$ is the superposition of the forces on $i$-th particle, from the trapping potential $U$ and interactions with other particles via WCA potential, i.e. ${\bf{F}}_i=-{\bf\nabla}_iU+\sum_j {\bf{F}}_{ij}^{wca}$.

While self-propelling in a fluid, an active particle not only produces force and energy but it also dissipates energy to the surrounding fluid. The dissipative force ($-\gamma{\bf{v}}_i$) acting on an $i$-th particle is proportional to its velocity ${\bf{v}}_i$ where the proportionality constant is the friction coefficient $\gamma$. 

Finally, if the active particles are moving in a fluid, it is expected that the fluid particles are also incessantly colliding with the active particle randomly (thermal collisions). The fluctuating force due to these collisions makes their positions and velocities even more noisy (note that their directions of motion were anyway stochastic by the random rotation implemented by ${\mathcal{R}}$). The thermal fluctuations are important in the case of microscopic active particles, such as active colloids \cite{Bechinger_Lowen_RMP}. This noisy force is given by $\sqrt{2\gamma K_BT}{\pmb\eta}_i(t)$ where $K_B$ is the Boltzmann constant, $T$ is the temperature of the fluid and ${\pmb{\eta}}_i$ is a random vector distributed with zero-mean, unit-variance Gaussian distribution. For larger active particles, thermal fluctuations are typically unimportant. However, there can be several other sources of athermal fluctuations due to complex interactions with the environment\cite{Bechinger_Lowen_RMP}.      

%Here ${i = 1,2,3...N}$  signify the particle-index. The number density of the particle fixed at $\phi={N\sigma^2}/{r_0^2}= 0.15$. 

Taking all the forces together, the equation of motion of  ${i}$-th particle can be represented within the Langevin paradigm \cite{gardiner1985handbook} as,
\begin{eqnarray}
m\frac{d{\bf{v}}_i}{dt}=-\gamma{\bf{v}}_i+{\bf F}_i +{\bf F}_i^{\text{active}}+\sqrt{2\gamma K_BT}{\pmb{\eta}}_i
\label{EquationOfMotion}
\end{eqnarray}

%%%%%%%%%%%%%%%%%%%%%%%%%%%%%%%%%%%

%where $m$ is the mass of a particle, $\gamma$ is the friction coefficient due to the surrounding fluid and ${\pmb{\eta}}_i$ is the thermal noise, having the following properties , $\langle{\pmb{\eta}}_i\rangle=0, \langle{\pmb{\eta}}_i(t_1).{\pmb{\eta}}_j(t_2)\rangle=\delta_{ij}\delta(t_1-t_2)$.  

The mass of the particle is considered to be unity and the friction due to the surrounding fluid on the particles, i.e. $\gamma$ is fixed at a high value such that for unit $\epsilon$ and $\sigma$, $\sqrt\epsilon/\gamma\sigma << 1$ which can be observed in many systems with high viscosity such as colloidal and bacterial suspensions.

By varying two parameters in the model of confined active particles as described before, a wetting-dewetting transition occurs at the interface between the active particles and the wall of the confinement, which is followed by an order-disorder transition, within the self-organised cluster of the active particles\cite{RanaSamArnabSoftMatter2019}. One of the parameters is the dimensionless variable $S$ defined before and the other one is Peclet Number which quantifies the ratio between the self-propulsion and the thermal diffusion. The ratio thus takes into account both the friction-limited active and thermal velocities.  It is defined as  
 \begin{eqnarray}
\text{Pe}=\frac{f_0}{\gamma\sqrt{K_BT}}.
\label{PecletEquation}
\end{eqnarray} 

where if ${f_0}\rightarrow 0$ the system represents a passive system equilibrated at temperature T. We will introduce here a mechanical quench exploiting these transitions. Hence, due to completeness, we here summarise the transitions detailed in \cite{RanaSamArnabSoftMatter2019} in the next section. 
\begin{figure*}%[!h]
   % \centering
   % \includegraphics[width=2\linewidth]{Figure1.pdf}
    \includegraphics[width=\textwidth]{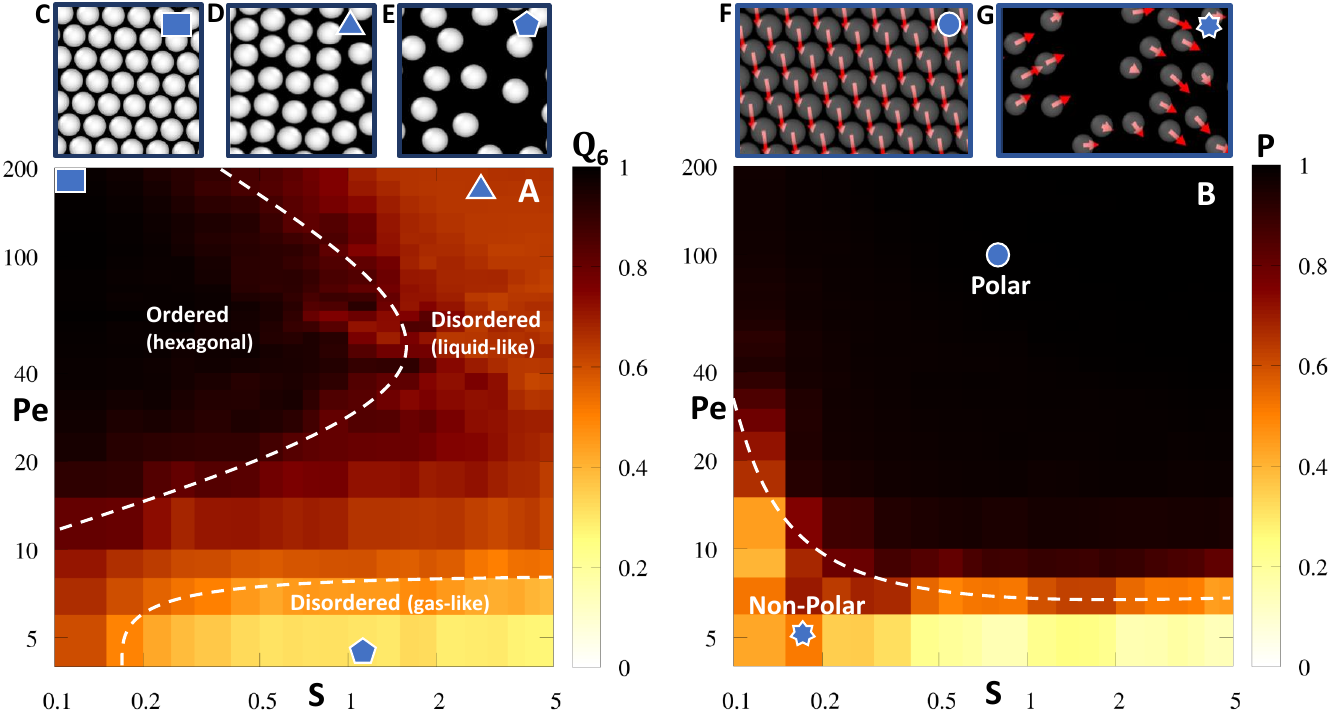}
    \caption{Phase Diagram: (A) depicts the phase diagram of distinct steady-state phases i.e. hexagonally ordered phase, liquid-like disordered phase, and gas-like disordered phase plotted in the $S$-Pe plane with the heat map representing $Q_6$ (i.e. Global Hexagonal Order Parameter). It is obtained after the system has reached a steady-state configuration from a randomly distributed configuration. In (A) the phases are separated by a schematic phase boundary (white broken lines). A typical configuration of hexagonally ordered phase is zoomed in panel (C) where the solid blue rectangle corresponds to the values of Pe and $S$ from panel (A). A typical liquid-like disordered phase is zoomed in the panel (D). The solid blue triangle correspond to the values of Pe and $S$ from panel (A). The gas-like phase is zoomed in the panel (E), and the solid blue pentagon correspond to values of Pe and $S$ from panel (A). The polar and non-polar non-equilibrium phases of the system in the $S$-Pe plane are plotted in panel (B) with the heat map representing the global polar order parameter $P$. The phases are separated by a schematic phase boundary (white broken line). The polar phase is zoomed in panel (F) where the arrow represents the velocity vectors of an individual active particle. The non-polar phase is zoomed in panel (G). The solid blue circle and star represent the corresponding values of Pe and $S$ from panel (B).}
    \label{fig:stiff}
\end{figure*}

\begin{figure*}%[!h]
    \includegraphics[width=\textwidth]{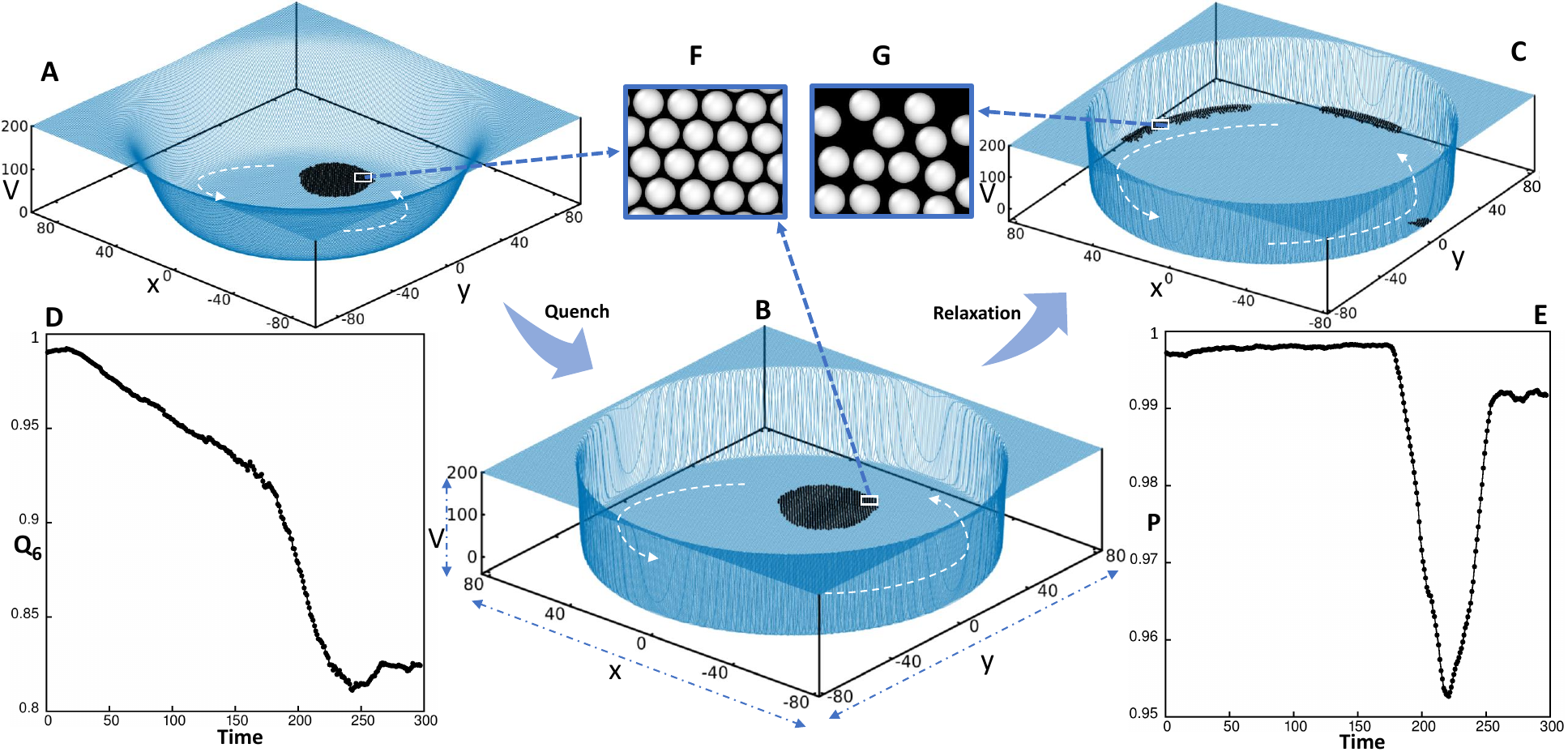}
    \caption{Mechanical quench: The steady-state non-equilibrium configuration of the system obtained for Pe$=200$ and $S=0.1$ is depicted in panel (A). The compact cluster formed with $N=1000$ active particles is represented in black with the trapping potential in blue. The zoomed configuration of the compact cluster is represented in panel F. The steady-state configuration of the cluster is obtained with Pe$=200$ and $S=0.1$.  Now, the steepness of the trap is instantaneously quenched to $S=2.5$ (B). However the particles take time to relax to a polar but liquid-like disordered state (C). The white dashed line with an arrow in panels (A),(B), and (C) represents the direction of rotation of the cluster along the boundary. The time variation of the global hexagonal order parameter after the quench is represented in panel (D). Similarly, the time variation of the global polar order parameter is shown in panel (E). }
    \label{fig:Quench}
\end{figure*}

\subsection{\label{sec:citeref}Boundary Driven Phase Transition}
The phase behavior and non-equilibrium phase transitions in active systems under various conditions are of current interest (for a review one may consult \cite{Ramaswamy_RMP, Bechinger_Lowen_RMP}). It can exhibit features that are fundamentally different from what we have in equilibrium systems. The model active system described before,  also exhibits a rich variety of phase behavior that has been discussed in detail by \cite{RanaSamArnabSoftMatter2019}. In \cite{RanaSamArnabSoftMatter2019}, it has been shown that for considerably high Pe and $S$, (i.e. when the potential is quite hard and steep), the confined active particles rotates along the circular wall of the potential, forming a thin wetting layer on it. The rotation can either be in clockwise or in counter-clockwise direction, chosen spontaneously by the particles. As the particles tend to follow each other via Eq.[\ref{activeforce_2}], once the choice has been made, they continue to rotate along the chosen direction. As all the particles in the layer are rotating along the circular boundary of the trap with little fluctuation (deviation from the mean angular velocity), it becomes a highly {\emph{polar}} layer. Its polarity is measured by the polar order parameter defined as,

\begin{eqnarray}
P=\langle|\frac{1}{N_i}\sum_{i}{\hat{n}_i}|\rangle
\label{PolarOrderParameterEquation}
\end{eqnarray}

where the angular brackets denote the average over all possible configurations and time both. Clearly, when all the particles move in the same angular direction then $P=1$, and if they move randomly then $P=0$. Thus, within a $0$ to $1$ scale $P$  measures the polarity of the system. Note that when the potential is steep,  the radially inward repulsive force from its wall acts only on those particles which are closest to the wall. This force balances the radially outward components of all the other forces associated with these particles. Only the angular components of the forces survive. This causes the particles to rotate along the circular wall of the trap in the low-noise regime. The rest of the particles, which are a little far from the steep wall of the boundary (and hence do not face it directly), simply follow the particles closest to the wall and start rotating with them. Hence the polar, wetting layer of the confined active particles forms and rotates along the steep, circular boundary around the centre of the trap.  In reference \cite{RanaSamArnabSoftMatter2019}, the behaviour of the system in the context of $P$, for a range of values of Pe and $S$ has been explored.  This phase diagram is reproduced in Fig.\ref{fig:stiff}. One may note in the phase diagram (Fig.\ref{fig:stiff}) that below a certain Pe, the particles cannot form the polar layer for any $S$. Furthermore, to obtain a polar layer in the lower values of $S$, a slightly more Pe is required. However, the dependence of $P$ on $S$ is weak.        
Apart from the orientational (here, polar) order, the positional order of the particles also undergo intricate phase transformation while varying $S$ and Pe. This phase diagram is also reproduced in Fig.\ref{fig:stiff} where the positional order is measured by the following hexatic order parameter

\begin{eqnarray}
&&\psi_{6,i}=\frac{1}{N_i}\sum_{j=1}^{N_i}e^{i6\theta_{ij}}\\ \nonumber
&& Q_6=\frac{1}{N}\sum_{i}\psi_{6,i}
\label{LocalHexagonalOrderParameterEquation}
\end{eqnarray}

Here $j$ is the index for $N_i$  neighbours of $i$-th particle, $\theta_{ij}$ is the bond angle between $i$-th and $j$-th particle with respect to an arbitrarily chosen reference frame, $\psi_{6,i}$ is the local hexatic order parameter and $Q_6$ is global hexatic order parameter.  In two dimensions, for a perfectly ordered triangular lattice structure $Q_6$ is $1$, for a perfectly disordered state it is $0$, and for the intermediate states $0 < Q_6 < 1$.  In Fig.\ref{fig:stiff}, the steady state configuration and time averaged $Q_6$ are computed for a range of $S$ and Pe to obtain the phase diagram. From the phase diagram, it is clear that at higher values of $S$, the particles are not in a positionally ordered state for any values of Pe. However, for smaller values of $S$ and a moderate range of Pe, hexatic order emerges within the clusters of the confined active particles. The emergent hexatic order at lower values of $S$ is preceded by a de-wetting transition where the wetting polar layer of active particles that rotates along the steep wall of the trap, de-wets upon decreasing the steepness of the wall to transform itself into a round cluster. The dynamics of the cluster is more intricate than the layer. Similar to the layer, it rotates along the trap boundary. However, while rotating, the cluster rolls about its center of mass due to the large (cluster)-scale torque generated within the cluster. It is the interplay among the active force, inertia of the particles, and the trap that generates large-scale torque within the cluster. When the cluster rolls and rotates simultaneously along the boundary of the trap, it gets compressed from all the directions by the potential. Hence, the positional (hexatic) order emerges within the cluster.

\subsection{\label{sec:citeref}Mechanical Quench}

Now we will mechanically quench the boundary of the trap. In particular, while the ordered cluster of the active particles are in steady rotating (about the centre of the circular trap ) as well as rolling (about the centre of the mass of the cluster) state being confined by the trap with soft boundary, we make the boundary very hard instantaneously. The cluster will collide with the hard boundary suddenly. Therefore hexatic as well as polar both the order decay sharply. Though the polar order recovers over time the hexatic order does not. This is consistent with the phase diagrams in Fig.\ref{fig:stiff}. According to the polar order phase diagram, above a certain threshold value of Pe, the polar order is stable for both soft as well as hard boundary.Therefore the polar order recovers after the drop due to the quench. However, the hexatically ordered state is unstable when the boundary of the trap is very hard. Hence it does not recover (see Fig.\ref{fig:Quench}).

The mechanical quench, as mentioned before,  is an event associated with the boundary. Intuitively, the particles close to the boundary will sense it first. Then the {\emph{information}} regarding the quench should flow throughout the cluster. Here our aim is to quantify the flow of the information by estimating the spatial as well as temporal extent of the information flow in our simple set up as described before. 

To quantify the information we need to define it. In the next section we will discuss the definitions. However, before going into it, it should be clear that after the quench takes place, the cluster suddenly encounters the hard boundary. Therefore, the particles close to the boundary undergo extreme deformation. As all the particles are interacting with each other, the deformation propagates with time through out the cluster. It re-distributes the velocity and position of all other particles eventually. Therefore, it is natural that the time dependent, joint position and velocity distribution of all the particles to carry the information regarding the quench.

\subsection{\label{sec:citeref}Information}
%\subsubsection{{\label{sec:citeref}Quenching of trap}}
 % We will

In this section we discuss definitions of different types of information. We consider a random process $X$ producing the random values $\{x_i\}$ where $i=1,2,...M$ where $M$ is an integer. $P(x_i)$ is the probability of obtaining the value $x_i$. As the least probable event has maximum information, one can define information corresponding to the outcome $x_i$ in bits as\cite{ThomasCover} $\mathcal{I}=\log_2\frac{1}{P(x_i)}$. The average (over all possible outcomes) information is given by $\langle\mathcal{I}\rangle=-\sum_iP(x_i)\log_2P(x_i)$, which is also known as Shannon entropy\cite{Shannon_1948,ThomasCover}. If the probability of an event is low, the surprise associated with it is high, and vice versa. Hence $1/P(x_i)$ can be interpreted as surprise associated with the event $x_i$.

Now we consider there are two such random processes $X$ and $Y$ producing random values $\{x_i\}$ and $\{y_i\}$ respectively where $i=1,2,...M$. The processes can be coupled to each other and in that case they are statistically dependent on each other, i.e. $P(x_i,y_i)\neq P(x_i)P(y_i)$ where $P(x_i,y_i)$ is the joint probability of obtaining $x_i$ from $X$ and $y_i$ from $Y$ simultaneously. The mutual information\cite{WicksPRE2007} that $X$ knows about $Y$ and $Y$ knows about $X$ is given by $I_{XY}=\sum_iP(x_i,y_i)\log_2\frac{P(x_i,y_i)}{P(x_i)P(y_i)}$. Note that $I_{XY}$ is zero (i.e. $X$ does not know about $Y$ and vice versa) when $X$ and $Y$ are two uncoupled, statistically independent process. Therefore, $I_{XY}$ is a measure of the dependence between two random processes. In other words it is a measure of the average information about $X$, which is communicated to $Y$ or vice versa. $I_{XY}$ can also be viewed as the Kullback–Leibler divergence\cite{KullbackLeiber1951} between $P(x,y)$ and $P(x)P(y)$ which is essentially interpreted as the {\emph {distance}} between these two distributions.    

From the definition $I_{XY}=I_{YX}$, i.e. the average information about $X$ communicated to $Y$ is always same as that of the other way round which also implies that mutual information is insensitive about the direction of the information flow. Mutual information does not tell us whether the information is transferred from $X\rightarrow Y$ or from $Y \rightarrow X$. It simply confirms us that X and Y share some information with each other.  However, in reality the direction of communication is often important.  Particularly in two cases: (1) where the information is supposed to flow from one place to another such that there is a built-in direction in the problem and (2) $X$ and $Y$ are non-reciprocally coupled i.e. the coupling strength between $X$ and $Y$ is not the same as $Y$ and $X$ . Hence the direction of communication between $X$ and $Y$  matters. Here we will be concerned only about the case (1). The interaction between $X$ and $Y$ are still reciprocal.

Now if we consider the random processes X and Y as time series, i.e. the respective random outcomes $x(t)$ and $y(t)$ are produced along the time axis  and we want to know the amount of directed information transferred between them without knowing their internal interactions then, according to \cite{SchreiberPRL2000} the relevant quantity is called global transfer entropy ($GTE$) defined as,

\begin{equation}
\begin{aligned}
{GTE}_{Y \rightarrow X}(t) &= \frac{1}{N}\sum_{i}\frac{1}{N_i(t)}\sum_{j}P(x_i(t+1),x_i(t),y_j(t)) \\
 &\times \log_{2}\frac{P(x_i(t+1)|x_i{(t)},y_j(t)}{P(x_i(t+1)|x_i(t))}. 
\end{aligned}
\label{TransferEntropyEquation}
\end{equation}
with a single step memory. Here $P(A|B)$ denotes the probability of occurring the event $A$ when the event $B$ has already occurred. The formula [\ref{TransferEntropyEquation}] can be generalised to any step memory. However here to keep our approach simple. We consider only one step memory. Note that the global transfer entropy respects the direction of the flow of information and therefore ${GTE}_{Y \rightarrow X}(t) \neq {GTE}_{X \rightarrow Y}(t)$.

Now we relate the dynamical variables of our problem and their probabilities with the random variables and probabilities used to define $GTE_{Y\rightarrow X}$ to compute it for our system before and after the quench.  Here, the index $i$ and $j$ are particle-index. The index $i$ is used for the {\emph{focal}}
particle towards which the information from its neighbours is communicated. The index $j$ represents the neighbours of $i$-th focal particle. $N_i(t)$ represents number of neighbours of $i$-th particle at time $t$. The neighbours of $i$-th focal particle at time $t$ are defined to be those particles which are within a circle of a certain cutoff radius surrounding the $i$-th particle. 

The random variable $x_i(t)$ represents the random $x$ and $y$ positions together with the velocity direction of $i$-th particle  at time $t$. Similarly the random variable $y_j(t)$ represents the random $x$ and $y$ positions together with the velocity direction of $j$-th neighbour of $i$-th particle at time $t$.  In the result section we will describe how this quantity captures the effect of the quench. However, as it contains a sum over all the particles, it cannot capture the spatial variation of information. In other words, it is unable to describe the flow of information.   

\begin{figure}%[!h]
    \includegraphics[width=8cm, height=6cm]{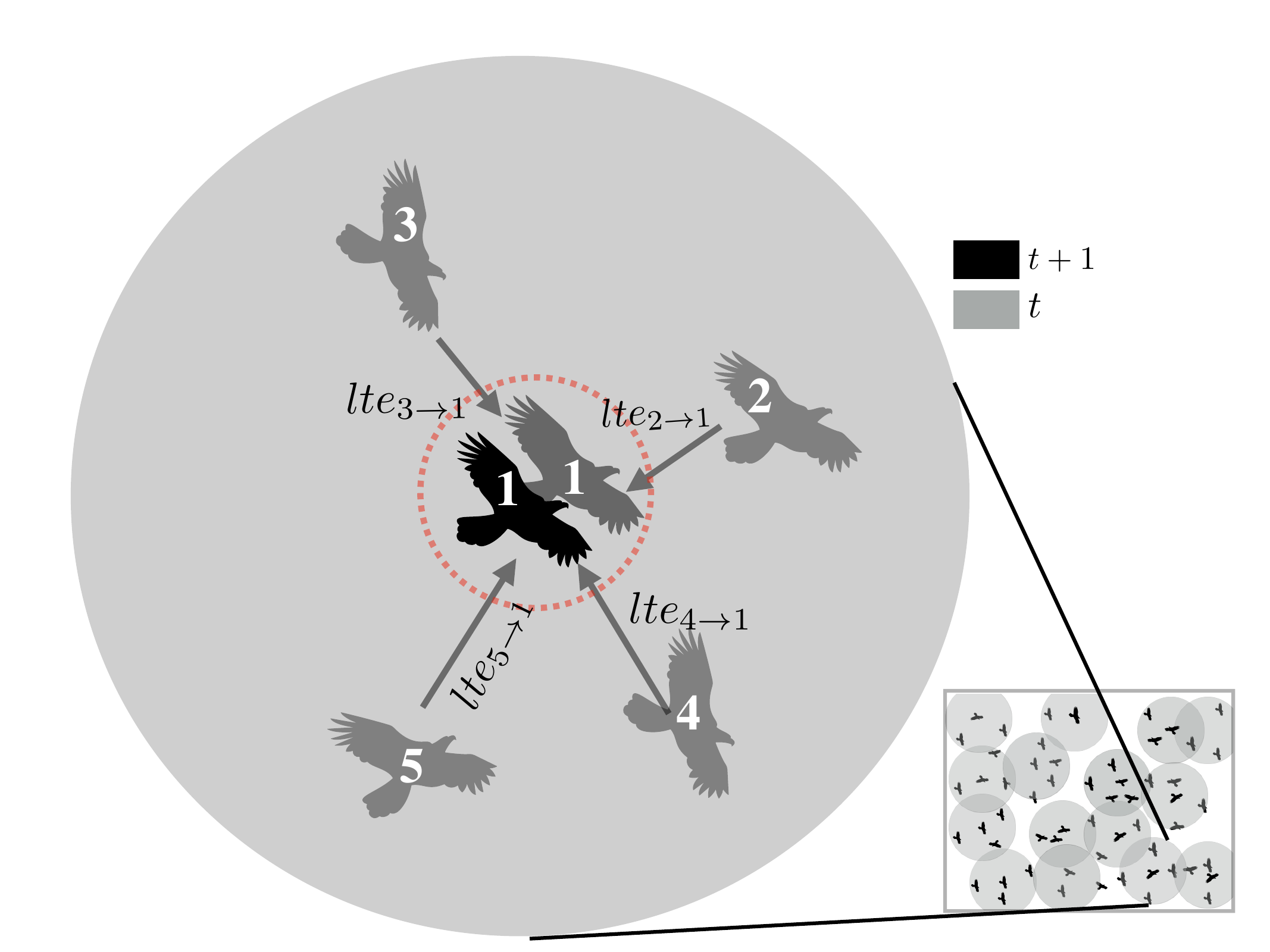}
    \caption{Schematic diagram illustrating local transfer entropy: In the right corner a schematic snap of self-propelling particles (birds) is shown where few focal particles are shown around which interaction circles having cut-off radius are drawn. Within the circles there are neighbours of the corresponding focal particles. Such a circle is zoomed in where (1) is the focal particle with (2)-to-(5) neighbours. (1) in grey obtains information from each of its neighbours at time $t$ and go to the next time step at $(t+1)$ denoted by black. By averaging the information over all 4 neighbours local transfer entropy as in Eq[\ref{LocalTransferEntropyEquation}] is obtained.}
    \label{figTE}
\end{figure}

To quantify the flow of the information we consider the average (over the number of neighbours) information (transfer entropy) obtained by $i$-th focal particle from its neighbours at time $t$ as,

\begin{equation}
\begin{aligned}
{LTE}_{Y \rightarrow X}(i,t) &=\frac{1}{N_i(t)}\sum_{j}P(x_i(t+1),x_i(t),y_j(t)) \\
 &\times \log_{2}\frac{P(x_i(t+1)|x_i{(t)},y_j(t)}{P(x_i(t+1)|x_i(t))}. 
\end{aligned}
\label{LocalTransferEntropyEquation}
\end{equation}

which can also be termed as local transfer entropy\cite{LizierPRE2008,LizierSpringerBook2016,LizierEPJB2010,ProkopenkoEntropy2013} ($LTE_{Y\rightarrow X}$). It is a space and time dependent quantity that can appropriately capture the flow of information occurring among the particles after the quench. The local transfer entropy is explained graphically in Fig.[\ref{figTE}]. In the result section, we will explain how the quantity describes the flow. Note that by definition $\frac{1}{N}\sum_i LTE_{Y\rightarrow X}(i,t)= GTE_{Y\rightarrow X}(t)$.  

One may note here that transfer entropy is a nonlinear generalization\cite{SchindlerAMS2011} of Granger causality\cite{Granger} (that tests causality in terms of prediction) that originated from information theory. It is a model-free estimate accounting for both linear and non-linear\cite{HLAVACKOVASCHINDLER20071} causal effects. The application of transfer entropy can be found in the dynamics of swarms \cite{WangPLOS2012}, neuroscience\cite{WibralFrontiers2015}\cite{WibralPBMP2010}\cite
{VicenteJCN2011}, robotics\cite{WirkuttisEntropy2023}, etc.  

In many complex systems, such as biological networks, the interactions between components are not uniform throughout the system but vary depending on local conditions. Therefore, studying local transfer entropy can help to reveal the underlying mechanisms driving interactions between components in specific regions of the system. For example, in the context of brain networks\cite{ParenteNatue2021}, local transfer entropy can be used to identify regions that are strongly connected and play a key role in information processing. Furthermore, local transfer entropy can provide a more detailed understanding of how information flows within a system\cite{WibralFrontiers2015}. By analyzing the transfer entropy between multiple pairs of variables in different regions of the system, researchers can identify the specific pathways through which information is transmitted. This information can be used to design interventions or control strategies to manipulate the system's behavior or to predict how the system will respond to perturbations. 

To estimate the probability mass function for the local transfer entropy we use the Java Information Dynamics Toolbox (JIDT) where the Kraskov-Stogbauer-Grassberger\cite{KSGPRE2004} estimator is implemented to calculate transfer entropy\cite{LizierFrontiers2014}.  

%\subsubsection{{\label{sec:citeref}Transfer Entropy}}
%If we consider the uncertainty measurement x of a random variable X with the probability function $p(x)$ defined over $\beta_{x}$ for $x \in \beta_{x} $. Then the Shannon entropy\cite{ThomasCover} which is the fundamental quantity of information theory, can be defined as 
%\begin{eqnarray}
%\text{H(X)}=-\sum_{x\in \beta_{x}}{p(x)}\log_{2}p(x).
%\label{ShanonEntropyEquation}
%\end{eqnarray} 
%By using the logarithm to base 2 the entropy measurement is in bits. The probability function should fall on the condition ${p(x) \geq{0}}$ and $\sum_{x\in \beta_{x}}{p(x)=1}$ . Hence Entropy can also be interpreted as a measure of average uncertainty\cite{ThomasCover}. Since surprise is, in some way inversely related to probability, when the probability of an event is low the surprise is high, and vice versa. Also if we know the entropy of a distribution then we can determine the number of bits that will be used to convey the information, in digital communication it is always desirable to communicate in the minimum possible bits so that we have higher efficiency. For a pair of random variables (X,Y) the joint distribution can be defined as ${H(X,Y)}=-\sum_{x\in \beta_{x}}\sum_{y\in \alpha_{y}}{p(x,y)}\log_{2}p(x,y)$. This concept can be generalized to any number of random variables given a multivariate distribution.
%In the previous section, we described the entropy of a random variable which was a measure of uncertainty, but if we ask the question of how to measure the difference between two distributions it is simply the relative entropy or Kullback Leiber distance\cite{KullbackLeiber1951} ${D(p||q)}=\sum_{x\in \beta_{x}}{p(x)}\log_{2}\frac{p(x)}{q(x)}.$ Note this is not a true distance as it is not symmetric under triangle inequality. Now what if by using one variable X we can infer some information about another variable Y, then the shared amount of information between X and Y is defined as ${I(X;Y)}=\sum_{x\in \beta_{x}}\sum_{x\in \alpha_{y}}{p(x,y)}\log_{2}\frac{p(x,y)}{p(x)p(y)}$ and is known as mutual information with joint probability distribution $p(x,y)$ it can also be represented as $I(X;Y)\equiv H(X) - H(X|Y) \equiv H(Y) - H(Y|X) \equiv H(X)+H(Y) - H(X,Y) \equiv H(X,Y) -H(X|Y) - H(Y|X)$, note that $I(X,Y)$ is symmetric under the exchange of $X$ and $Y$ and also we can express mutual information with entropy of a random variable with itself as $I(X;X)=H(X)-H(X|X) = H(X)$. Mutual information does not tell us if the information is transferred from $X\rightarrow Y$ or from $Y \rightarrow X$ it simply tells us that X and Y share some information with each other.  

%Now if we have two time series variables X and Y and we want to know the amount of directed information transferred between them without knowing their internal interactions then Schreiber’s\cite{SchreiberPRL2000} criteria requiring the deﬁnition to be both directional and dynamic is defined to  be 

\subsection{\label{sec:citeref}Results}

\begin{figure*}%[!h]
    \includegraphics[width=\textwidth,height=4cm]{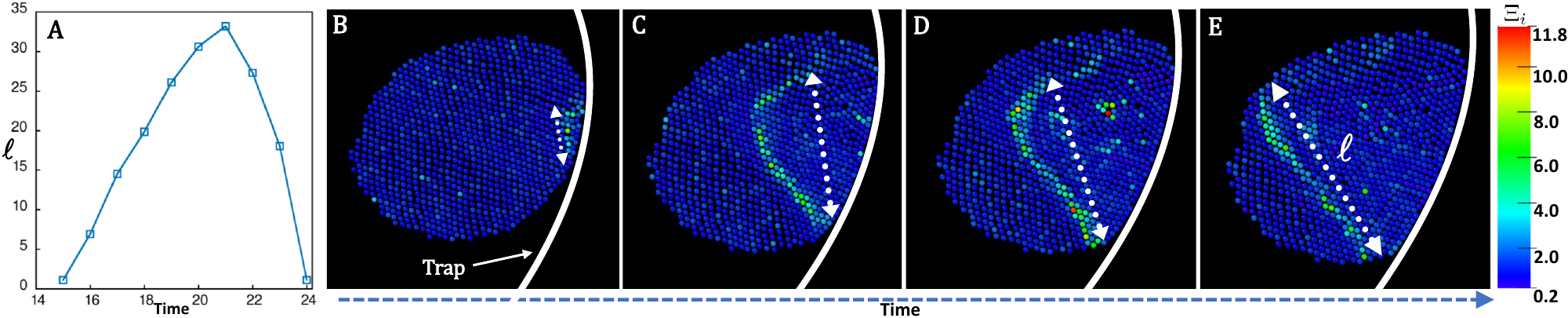}
    \caption{Local Transfer Entropy front : The information propagates within the cluster after the quench with time in the form of a front whose size increases up to the linear dimension of the cluster and then decreases with time. The length of the front is denoted by $l$, which is measured  as the distance between the two furthermost particles having large (> 4) transfer entropy, and is depicted in panel (A). The increase of the front-length with time is depicted in panel (B-E) where the trap boundary is represented in white, and the color represents the transfer entropy values.}
    \label{fig_Length Scale}
\end{figure*}

\begin{figure*}%[!h]
\includegraphics[width=\textwidth,height=13cm]{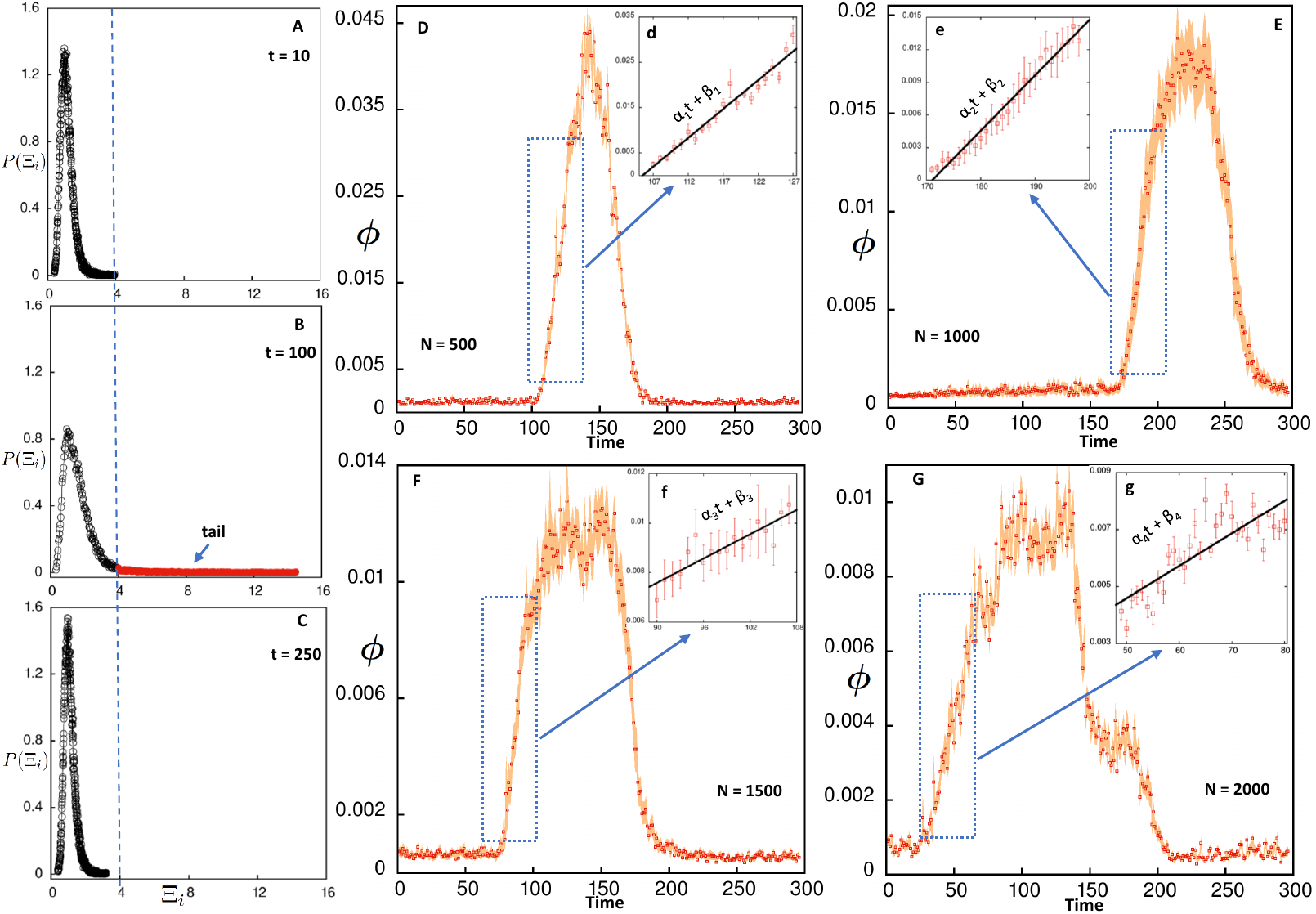}
    \caption{Probability density snapshots at different time points: The distributions of $LTE_{Y\rightarrow X}$  are plotted in panel (A), (B) and (C) for three different time point (shown in the corresponding plots). Panel (A) depicts the probability density distribution before the quench takes place, panel (B) shows the probability distribution after the quench takes place and panel (C) shows the probability distribution long after the quench. The blue dotted line is for $LTE_{Y\rightarrow X}=4$ which is an arbitrarily chosen threshold that determines the tail of the distribution shown in panel (B), colored in red. The tail is significantly small for panels (A) and (C).  Tail area: The area under the tail is represented by $\phi$.  The time evolution of the tail area is plotted in panel (D)-to-(G) for various system sizes (N=500,1000,1500,2000) and averaged over 20 configurations for each. The standard deviation is represented in yellow. The growth in the tail area (denoted by the blue rectangle) is best-fitted with the form $\alpha_i t + \beta_i $ and is shown in the inset panels (d)-to-(g) and drawn in solid black lines with the standard deviations in red square. The best-fit values of $\alpha_1 = 0.00127, \beta_1= 0.13398 $, $\alpha_2 = 0.00051, \beta_2 = 0.08740$,$\alpha_3 = 0.00016 , \beta_3= 0.00717$,$\alpha_4 = 0.00011, \beta_4= 0.00110$ .}
    % $\alpha_1 = 0.00127039, \beta_1= 0.13398 $, $\alpha_2 = 0.000510792, \beta_2 = 0.0874008$,$\alpha_3 = 0.000163886 , \beta_3= 0.00717$,$\alpha_4 = 0.000113767, \beta_4= 0.00110196$ .
    \label{fig:Tail area}
\end{figure*}

\begin{figure}%[!h]
   % \centering
   % \includegraphics[width=2\linewidth]{Figure1.pdf}
    \includegraphics[width=8.5cm, height=11.5cm]{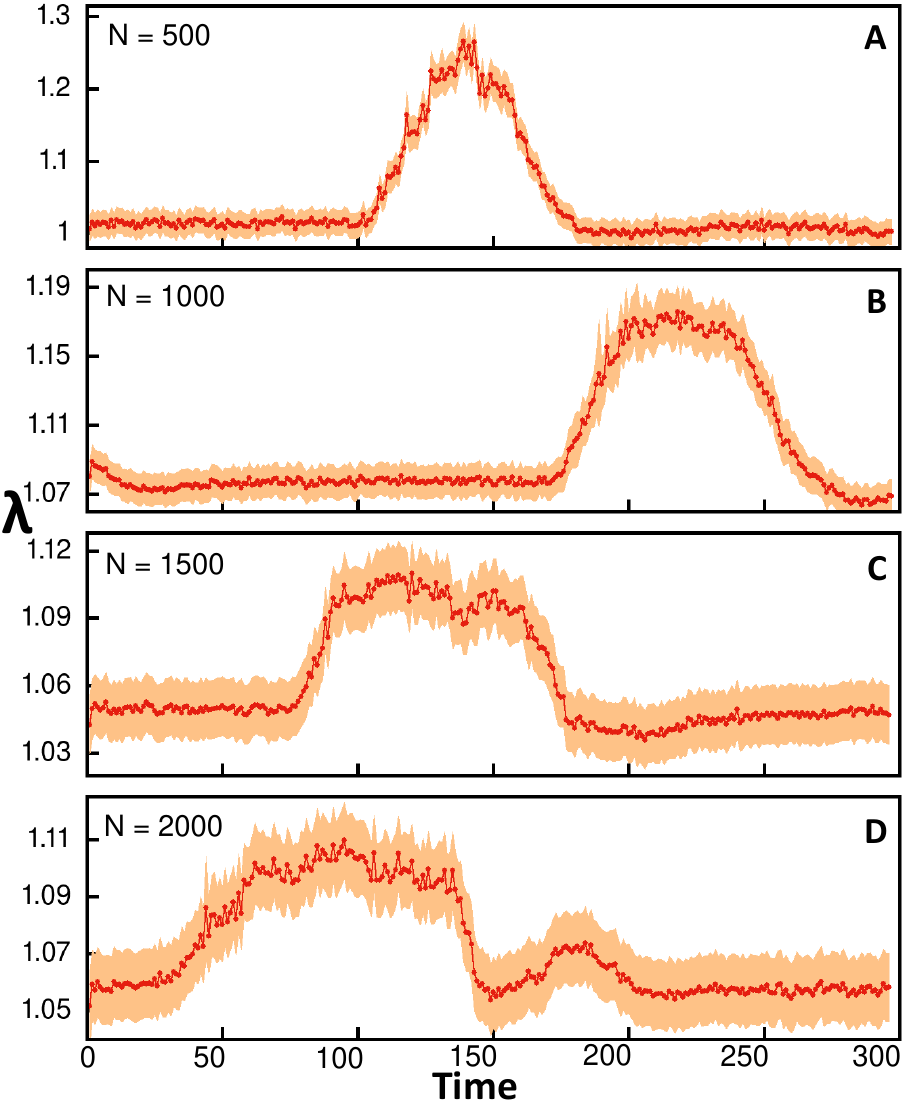}
    \caption{Global Transfer Entropy vs Time. The four different panels (A-D) depict the $GTE_{Y\rightarrow X}$ in the y-axis versus time in the x-axis for different cluster sizes (N=500 for panel A, N=1000 for panel B, N=1500 for panel C, N=2000 for panel D) averaged over 20 realisations each. The standard deviation is colored in yellow. Clearly, we see that there is a hump in all the four plots suggesting an increase in information flow after the quench.}
    \label{fig_GTE_vs_time}
\end{figure}

Here, we will explain how the local transfer entropy is calculated and from there how one can extract important features of the flow of information. The local transfer entropy is calculated from the probabilities involved in Eq.[\ref{LocalTransferEntropyEquation}]. These probabilities are calculated from the configurations (i.e. velocities and positions of the particles) of the system at two consecutive times $t$ and $t+1$. The configurations are obtained by simulating the system i.e. integrating the equation of motion of individual particles given in Eq.[\ref{EquationOfMotion}]. While calculating the local transfer entropy for a given particle, one has to consider the particle to be a focal particle (say, $i$-th focal particle) towards which the information is communicated from its neighbours (say, $j$-th particle) at a given time. The neighbours of a focal particle in a given configuration are those particles that are inside a circle of the cut-off radius centering the focal particle. Inside the circle, in order to calculate the local transfer entropy of an $i$-th focal particle, first one need to calculate the summand in Eq.[\ref{LocalTransferEntropyEquation}] for every $i$-$j$ pair and then the average of this quantity for all the $j$ neighbours has to be taken. The local transfer entropy of $i$-th focal particle, as calculated here, is considered to be the average information transferred to the focal particle from its neighbours at a given time point $t$.  We have used JIDT to calculate the local transfer entropy.    

First we analyse the local transfer entropy data simply by plotting the positions of the particles together with their respective local transfer entropy in colour scale (see Fig.[\ref{fig_Length Scale}]). The configurations used for the plot is taken from the instant at which the quench took place and afterwards.  From the snaps of the configurations shown in Fig[\ref{fig_Length Scale}] it is apparent that as soon as the cluster hits the hardened (due to the quench) boundary of the trap, only a small number of particles facing the boundary have large values ($>$ $4.0$, a chosen threshold value of $LTE_{Y\rightarrow X}$) of local transfer entropy $LTE_{Y\rightarrow X}$. These particles are localised in a small area close to the trap boundary.  The number of the particles having a large value of  $LTE_{Y\rightarrow X}$ increases over time and the region where these particles are localised in the cluster of the particles, moves like a front from the boundary towards the hindmost end of the cluster (see Fig[\ref{fig_Length Scale}]). It can be called as the information front that propagates. The number of the particles having a large value of $LTE_{Y\rightarrow X}$ ( > 4.0) is also plotted with time starting from the instant of the quench in Fig.[\ref{fig_Length Scale}]. It increases till the two ends of the information front reaches up to the boundary of the cluster. Then it starts to decrease simply because the cluster is a finite one. Eventually the size of the information front decreases and becomes zero as it propagates towards the end of the cluster. While measuring the length of the information front by counting the particles in it, we find that its maximum size (for all time steps ) is proportional to the linear size of the cluster i.e. $L \propto \sqrt{Area_{cluster}}$.  This is true for different cluster sizes, that is, for N = 500,1000,1500 and 2000.  Note that the area of the cluster is proportional to the number of particles present in the cluster. If all the particles present in the system form a single cluster then it is proportional to $N$. So, as the number of particles increases in a cluster, the size of the information front also increases and hence, eventually all the particles in the cluster will know about the perturbation i.e. the quench.  Note that the total time taken by the information front to originate after the quench, then to propagate and grow and finally shrinks to zero, is quite small in comparison to the time taken by the cluster to deform considerably due to the quench. Therefore, the bottom line of our present study is : (1) for a given parameter space where the underlying cluster of the active particles is well-ordered, the information i.e. the local transfer entropy flows not in a scattered way, instead it flows in a coherent manner forming a propagating front, and (2) the maximum size of the front is proportional to $N^{1/2}$ where $N$ is the total number of the particles.

Next we calculate the probability distribution of  $LTE_{Y\rightarrow X}$  over the particles for different times including the instant at which the quench took place. Three such distributions (one at pre-quench time, one just after the quench, and another long after the quench) are shown in Fig.[\ref{fig:Tail area}]. In the pre-quench distributions as well as in the distributions long after the quench, there are no largely deviated (from mean) values of $LTE_{Y\rightarrow X}$ whereas in the distributions shortly after the quench there are.  This is consistent because as we have discussed previously that after the quench takes place the information front appears, it grows and finally shrinks and disappears. Within the front, the particles have high values of local transfer entropy. This is reflected in the tails of the  $LTE_{Y\rightarrow X}$  distributions. Before and long after the quench when there is no particles having large value (say $>$ $4$, a chosen threshold) of  $LTE_{Y\rightarrow X}$,  the tails in the distributions are absent. Just after the quench, the largely deviated values starts to appear and therefore the tail in the distribution also appears. Over a certain time window after the quench, more and more particles obtain large values of  $LTE_{Y\rightarrow X}$, and therefore the tail grows. After reaching its maximum the number of particles having large values of local transfer entropy starts to decrease. Therefore the tails in the distributions also shrinks.  Finally, the information front of the particles having a large value of local transfer entropy disappears together with the tails of its distribution. Here, from the time evolution of the tail, in particular from the growth of the tail, we will find how the information front grows and propagates simultaneously in the cluster. For that, we will plot the fraction of the number of particles within the tail of the local transfer entropy distributions with time for four different $N$ (see Fig.[\ref{fig:Tail area}]). Before quench takes place it was very small, close to zero. From the instant of the quench, it increases and reaches up to a maximum after which it decreases to zero again. Our analysis (as shown in Fig.[\ref{fig:Tail area}]) confirms that from the instant when the quench takes place till the fraction reaches its maximum, the fraction of the number of the particles in the tail of the distributions (say, $n$) grows ballistically i.e.  $n\propto t$. This also implies that the rate of growth in the number of particles having large values of  $LTE_{Y\rightarrow X}$ is constant.  Hence we conclude that for the given set of parameters for which the cluster of the active particles is well-ordered, the information propagation through the cluster is ballistic.  

After exploring the spatio-temporal features of information propagation within an ordered cluster of active particles from local transfer entropy, finally we compute global transfer entropy ($GTE_{Y\rightarrow X}$) by taking the average of  $LTE_{Y\rightarrow X}$ over all the particles for different values of $N$ (see Fig.[\ref{fig_GTE_vs_time}]). From the time evolution of $GTE_{Y\rightarrow X}$ in Fig.[\ref{fig_GTE_vs_time}] it is apparent that it grows after the quench takes place, reaches to a maxima, and as the information front propagation completes over time,  $GTE_{Y\rightarrow X}$ decays and comes back to its minima.  The hump in the time evolution of the global transfer entropy captures the impression of the flow of information after the quench takes place. However, it cannot probe further to explore the spatio-temporal features of the flow, as done by computing local transfer entropy before.   

\section{Conclusion}

In this work, we have considered self-propelling as well as self-aligning particles moving in two dimensions being confined in a circular trap. The self-organisation of the confined particles crucially depends on the softness/hardness of the trap boundary. If the boundary is soft enough and the particles are moderately active, they form round-shaped clusters that rotate about the trap centre along its boundary and roll about its centre of the mass simultaneously. Such an intricate dynamics facilitates to establish hexatic order within the clusters on top of its pre-established polar order. However if the boundary is hard, the cluster stops rolling about its centre of mass. It only rotates along the trap boundary. The round shape of the cluster and its hexatic order becomes unstable when the trap wall becomes hard. Within the circular trap with a hard and steep boundary, the particles forms a layer that rotates along the boundary but does not roll. Hence the hexatic order becomes unstable. However, the layer of the particles still remains polar. The phase diagram of the system in the plane of the activity and softness (or, hardness) of the trap boundary, clearly manifests different phases for different sets of parameters of the system. 

Using the aforementioned non-equilibrium phase transformation driven by the boundary, a mechanical quench is introduced.  It makes the soft trap boundary a hard one instantaneously. Hence the round-shaped hexatic cluster relaxes to the polar but positionally disordered layer rotating along the hard trap boundary. However, together with the structural relaxation, we have shown here that the information regarding the quench flows throughout the cluster. It is a much faster process than the structural relaxation. We quantify the information by computing the local transfer entropy. We have shown that within the cluster, the information flows coherently as a front. It propagates ballistically from the particles close to the boundary towards the far end of the cluster. We have also shown that the maximum size of the front becomes proportional to the square-root of the number of particles present in the cluster. 

%So far the flow of the information is calculated only for a given Pe and for a pair of $S$ used for the quench. It will be interesting to explore the characteristic features of the information flow for various other set of parameters and find the correlations between the flow of the information and mechanics of the system. Furthermore, it will also be interesting to unfold the features of the information flow in the presence of non-reciprocal inter-particle interaction.

Currently, the information flow has been computed solely for a specific pair of Pe and $S$ used for the quench. However, it opens up several other important issues. For example, here the study is around the quench from the soft to hard confinement such that the information flows through a well-ordered (not perfectly though) structure. However, it is natural as well as important to ask what happens to the information flow if the quench is in the opposite direction (hard to soft). Note that in this case, the information will start to flow through a positionally disordered structure. Furthermore, one may also ask, instead of quenching the system mechanically, what happens to the information flow if the system is quenched via the strengths of the fluctuations present in the system. Additionally, uncovering the characteristics of the information flow in the context of non-reciprocal inter-particle interaction would also be of great interest. It would be intriguing to investigate such issues to explore distinctive attributes of the information flow for different sets of parameters and discern correlations between the dynamics of the information and the mechanics of the system.

\begin{acknowledgments}
AS acknowledges the Start-up Grant from UGC, and the Core Research Grant (CRG/2019/001492) from DST, Government of India. Md. Samsuzzaman (MS) extends thanks to Savitribai Phule Pune University for the JRD TATA and BB Laud fellowship. Mohammad Hasanuzzaman acknowledges the support received from the ICTP-TRIL Fellowship.

\end{acknowledgments}

\nocite{*}

\bibliography{apssamp}% Produces the bibliography via BibTeX.

\end{document}